\documentclass[a4paper]{jpconf}

\usepackage{citesort}

\newcommand{\be}{\begin{equation}}
\newcommand{\ee}{\end{equation}}

\newcommand{\bea}{\begin{eqnarray}}
\newcommand{\eea}{\end{eqnarray}}

\usepackage{graphicx}

\begin{document}

\title{Recent scaling properties of Bitcoin price returns}

%\noindent \qquad \\[-6pt] \qquad Version 2.0\\\qquad December 21, 2006

\author{T~Takaishi}

\address{Hiroshima University of Economics, Hiroshima 731-0192, JAPAN
}

\ead{tt-taka@hue.ac.jp}

\begin{abstract}
While relevant stylized facts are observed for Bitcoin markets,
we find a distinct property for the scaling behavior of the cumulative return distribution.
For various assets, the tail index $\mu$ of the cumulative return distribution exhibits $\mu \approx 3$,
which is referred to as "the inverse cubic law." On the other hand, that of the Bitcoin return
is claimed to be $\mu \approx 2$, which is known as "the inverse square law."
We investigate the scaling properties using recent Bitcoin data and
find that the tail index changes to $\mu \approx 3$, which is consistent with the inverse cubic law.
This suggests that some properties of the Bitcoin market could vary over time. 
We also investigate the autocorrelation of absolute returns and find that it is described by
a power-law with two scaling exponents.
By analyzing the absolute returns standardized by the realized volatility,
we verify that the Bitcoin return time series is consistent with
normal random variables with time-varying volatility.
\end{abstract}

\section{Introduction}

In 2008, Nakamoto\cite{Nakamoto2008} proposed Bitcoin as the first practical cryptocurrency based on the blockchain technology.
His proposal was quickly accepted, and in 2009, the Bitcoin network was launched as a peer-to-peer payment network. 
Since then, a number of digitized cryptocurrencies have been proposed and created\footnote{ e.g., see https://coinmarketcap.com/ for the current market capitalizations of cryptocurrencies.}, 
and growing cryptocurrency markets have a strong impact 
on contemporary financial markets. 
Bitcoin has recently attracted the interest of researchers and
various properties and aspects of Bitcoin have been investigated. For example, 
volatility analysis \cite{bouoiyour2016bitcoin,dyhrberg2016bitcoin}, 
price clustering\cite{urquhart2017price,li2018price},
adaptive market hypothesis \cite{khuntia2018adaptive},
transaction activity \cite{koutmos2018bitcoin}, 
multifractality\cite{takaishi2018},
liquidity and efficiency\cite{wei2018liquidity}, 
Taylor effect\cite{takaishi2018taylor},
structural breaks \cite{thies2018bayesian},
long memory effects\cite{phillip2019long},
rough volatility\cite{takaishi2019rough},
power-law cross-correlation\cite{takaishi2020power},
asymmetric multifractal analysis\cite{kristjanpoller2020cryptocurrencies},
and so forth.

Universal properties across various financial time series are classified as stylized facts\cite{Cont2001QF}, 
which include 
(i) fat-tailed distributions, (ii) volatility clustering, and
(iii) slow decay of autocorrelation in absolute returns, among others.
While stylized facts are also observed in Bitcoin, for example, \cite{chu2017garch,takaishi2018},
a distinct property has been observed for the tail index of return distributions.
For other assets such as stocks, it is known that the tail of the cumulative return distribution is described 
by a power-law function and its tail index $\mu$ is obtained as $\mu \approx 3$, known as "the inverse cubic law"\cite{gopikrishnan1998inverse,gopikrishnan1999scaling,plerou1999scaling,pan2007self}.
For the Bitcoin market before 2014, 
the tail index is estimated to be $\mu \approx 2$, which differs from the known stylized fact for other assets and 
is referred to as "the inverse square law"\cite{easwaran2015bitcoin}.
 Similar tail indices have also been reported in Ref.\cite{beguvsic2018scaling}.
Since this is a newly emerging market, 
market properties of Bitcoin could change as trading activity increases over time.
In fact, the market efficiency measured by the Hurst exponent of the return time series 
varies over time and at the early stage of the Bitcoin market,
the Hurst exponent is observed to be less than 1/2, which indicates that the time series is anti-persistent\cite{urquhart2016inefficiency}. 
This anti-persistent behavior could be related to the illiquidity of the Bitcoin market\cite{wei2018liquidity,takaishi2019market}.
As the liquidity of the Bitcoin market improves, the Hurst exponent moves to 1/2\cite{urquhart2016inefficiency,takaishi2019market}. 

We investigate the possible change in scaling properties of the return distribution for the recent Bitcoin market.
We also investigate the cumulative distributions and autocorrelation of the absolute returns and the return standardized by the realized volatility.

\section{Data set}
The data are Bitcoin tick data (in dollars) traded on Bitstamp 
from September 11, 2011 to June 21, 2020\footnote{Due to a hacking incident, 
no data are available from January 4, 2015 to January 9, 2015. For these missing data, 
we treat them as the price is unchanged.} and downloaded from Bitcoincharts\footnote{http://api.bitcoincharts.com/v1/csv/}.
We construct 1-min price data from the Bitcoin tick data and then calculate 1-min returns $R_t, t=1,2,...,N$ as
\be
R_t= \log p_{t} -\log p_{t-1},
\ee
where $p_t$ is the 1-min price at $t$ (min).
We further standardize the returns data using
\be
{\bar{R_t}} = \frac{R_t-\nu}{\sigma}. 
\ee
where $\nu$ and $\sigma$ represent the average and standard deviation of returns $R_t$, respectively.
We select the following two data periods from the data, which separate low and high liquidity periods: I) September 11, 2011-December 31, 2013 and 
II) January 1, 2015-June 21, 2020.
 Dataset I (II) contains low (high) liquidity data\cite{takaishi2019market}.

\section{Results}
First, we show the cumulative distributions of returns in Figure 1(a): positive tail and (b): negative tail, in a log-log plot,
and recognize that the tail behavior is described by a power law.
While there is no significant difference between positive and negative tails,
a considerable difference is seen between data periods, that is, a heavier tail is observed for the cumulative distribution of dataset (II),
which contains high liquidity data. 
We obtain tail indices of the cumulative distributions by a regression fit to a power-law function of $\kappa \bar{R}^{-\mu}$, where $\kappa$ and $\mu$ are 
fitting parameters. The results of the tail indices are reported in Table 1.
The tail index for dataset I is $\mu\approx 2$, which is consistent with the previous results obtained for earlier periods\cite{easwaran2015bitcoin}.
On the other hand, the tail index for II is $\mu \approx 3$, which is consistent with the tail index obtained for other assets such as stocks.
For the entire data set, we find a tail index between I and II. 
Our findings suggest that the scaling properties of the return distributions of the Bitcoin market could change over time, 
 For the recent liquidity of the Bitcoin market, the tail index is in agreement with the well-known stylized facts for other assets. 

\begin{table}
\centering
\caption{Results of tail index.}
%\scriptsize
\begin{tabular}{cccc}
\hline
                  & (I) 2011-2013 & (II) 2015-2020 & 2011-2020    \\
\hline
Fitting region      & [2,20] & [2,20] & [1,10] \\
positive tail index $\mu$      & -2.06 & -3.33  & -2.35 \\
negative tail index $\mu$      & -2.08 & -3.23  & -2.38 \\
\hline
\end{tabular}
%\vspace{-8mm}
\end{table}

\begin{figure}
%\vspace{5mm}
\centering
\includegraphics[height=6.5cm]{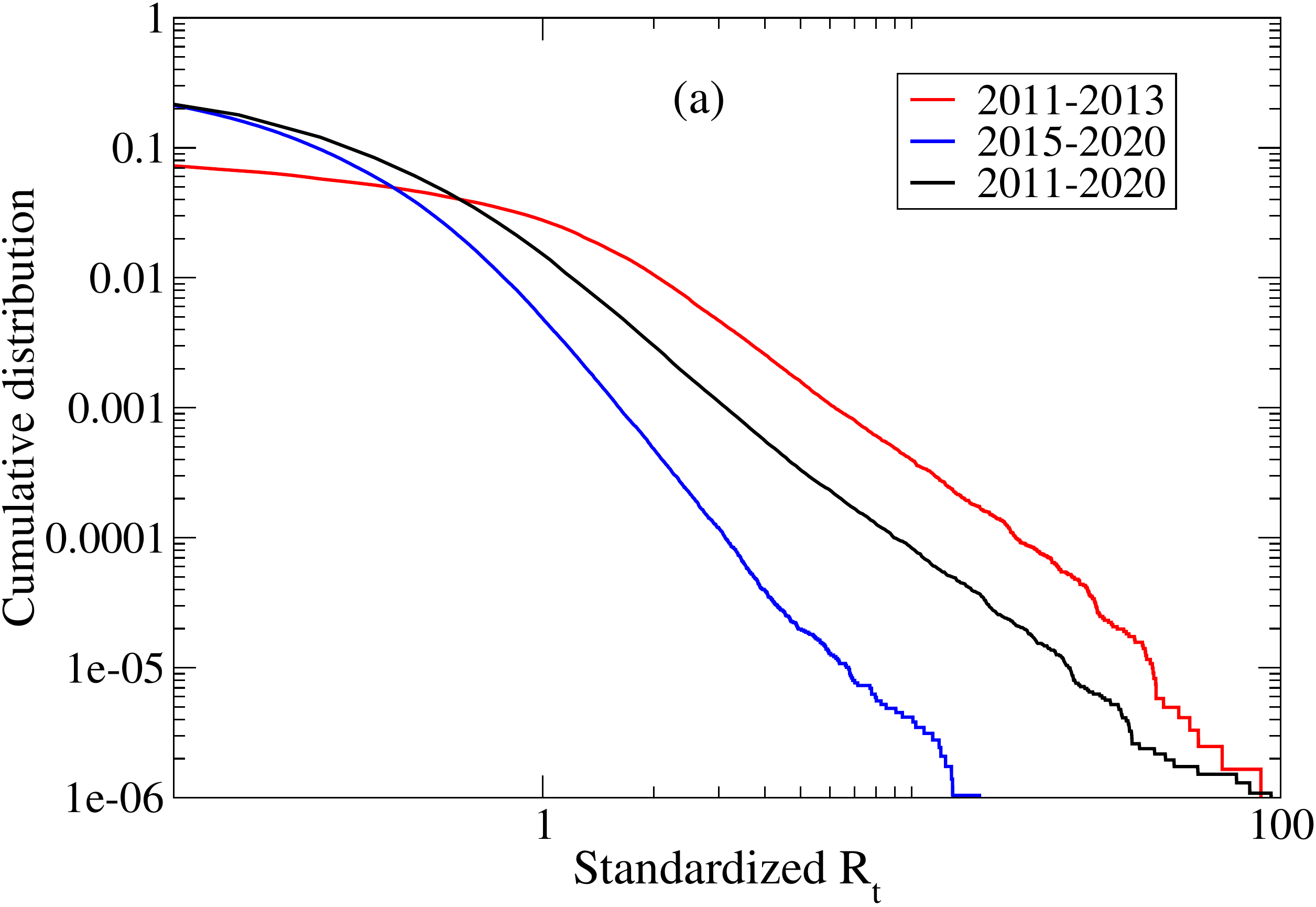}
\includegraphics[height=6.5cm]{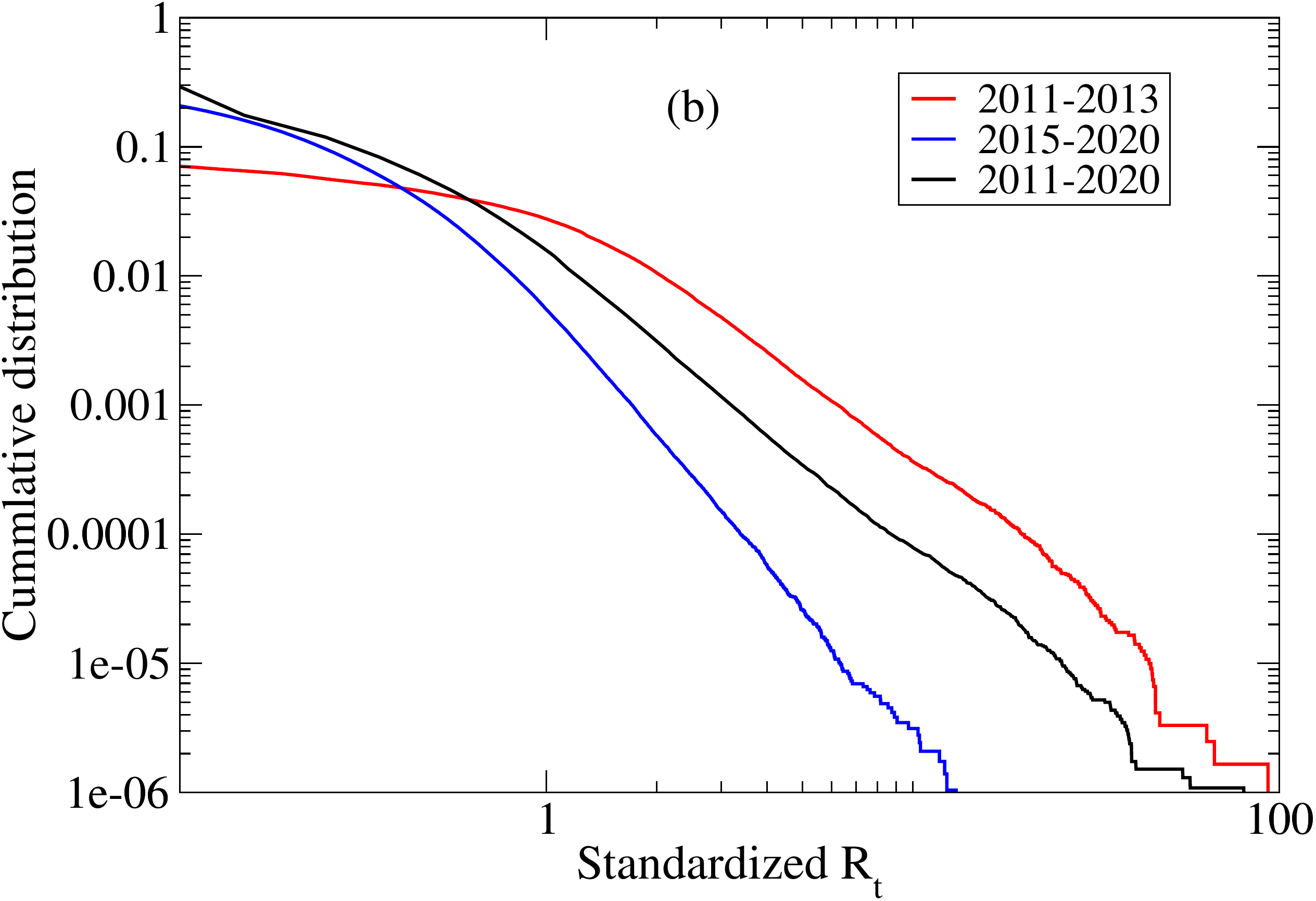}
\caption{
(a) Cumulative distribution for positive tail.
(b) Cumulative distribution for negative tail.
}
%\vspace{-2mm}
\end{figure}

Next, we show the autocorrelation function (ACF) of the absolute returns. 
The ACF of a time series $x_i:i=1,...,N$ is defined by
\be
ACF(t)= \frac{\langle(x_i-\rho)(x_{i+j}-\rho)\rangle}{\sigma_x^2}. 
\ee
where $\rho$ and $\sigma_x^2$ represent the average and variance of $x_i$, respectively, and
$\langle O_i \rangle$ means taking an average over $O_i$. 
Figure 2 displays the ACF of the absolute returns in the log-log plot and 
we see that the ACF is also well described by a power law. 
A remarkable characteristic of the ACF is that the power-law exponent seems to change at $t\sim 1000$.
The power-law exponent for $t>1000$ is found to be larger than that for $t< 1000$.
There is no clear explanation for this change.
The results of the power-law exponents are reported in Table 2. Furthermore, we find that there is no substantial difference for the power-law exponents between the three data sets. 
Thus, the long-memory property of Bitcoin remains unchanged over time.

\begin{table}
\centering
\caption{
Results of the power-law exponent.}
%\scriptsize
\begin{tabular}{cccccc}
\hline
                  & (I) 2011-2013 & (II) 2015-2020 & 2011-2020    \\
\hline
fitting region      & [3,500] & [3,500] & [3,500] \\
exponent $\mu$      & -0.121 & -0.116  & -0.120 \\
\hline
fitting region      & [15000,10000] & [1500,10000] & [1500,10000] \\
exponent $\mu$      & -0.197 & -0.235  & -0.206 \\
\hline
\end{tabular}
%\vspace{-8mm}
\end{table}

\begin{figure}
%\vspace{5mm}
\centering
\includegraphics[height=6.5cm]{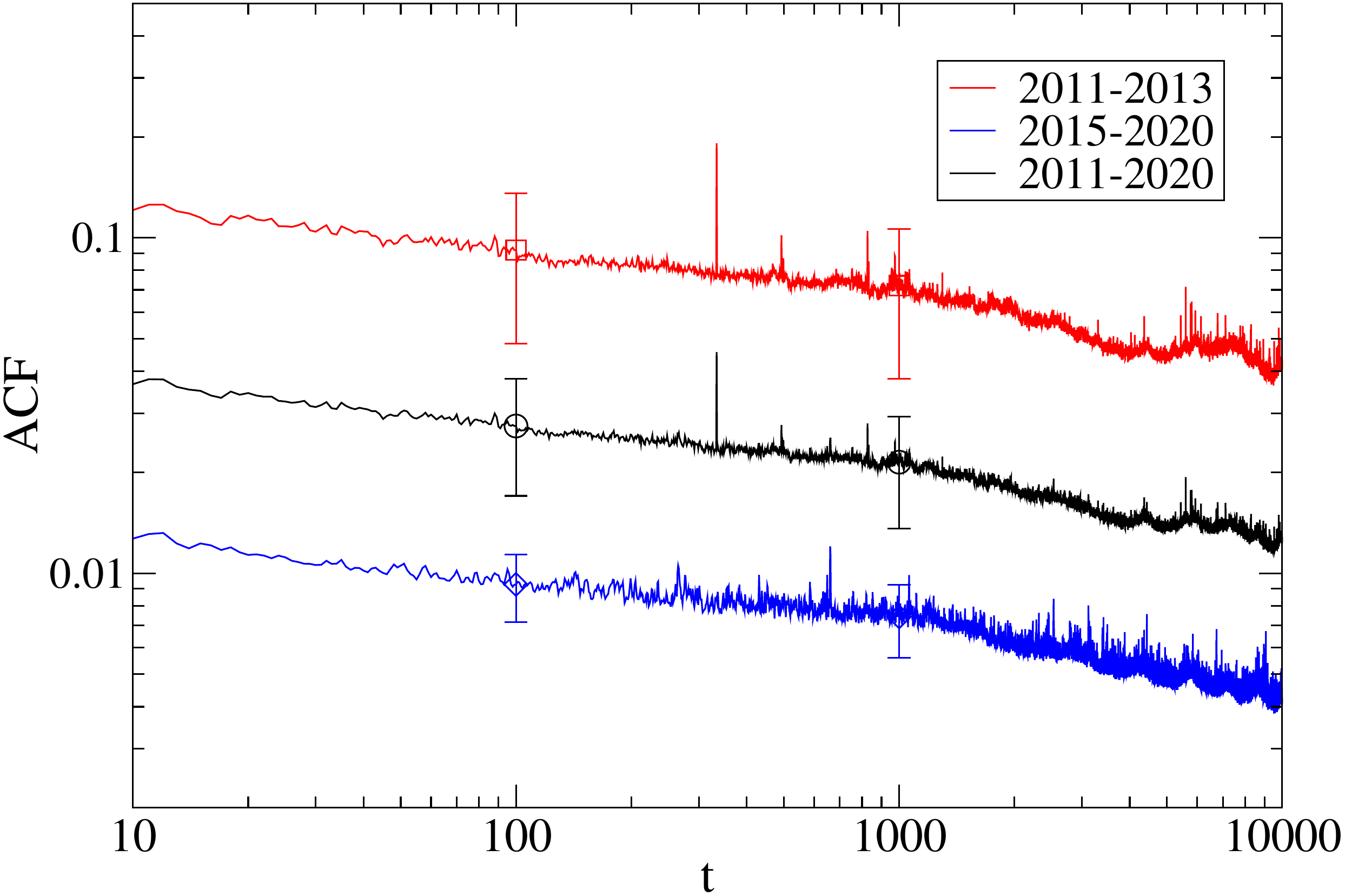}
\caption{
 ACF of absolute returns. The symbols at t=100 and 1000 show typical one-sigma errors calculated using the jackknife method.
}
%\vspace{-2mm}
\end{figure}

The power law observed in the absolute return means that the time series of the absolute return has long memory, which sharply contrasts to returns for which
the ACF of the return time series is short ranged.
The long memory in the absolute returns is considered to be related to the time correlation of volatility.
Let us assume that the return time series is described by $R_t \equiv \sigma_t \epsilon_t$, where $\sigma_t$ is the volatility at $t$ and 
$\epsilon_t$ is the standard normal random variable. Namely, the return $R_t$ is assumed to behave as a normal random variable with time-varying volatility.
This assumption of returns automatically satisfies no autocorrelation in returns.
On the other hand, the autocorrelation of absolute returns depends on the time structure of $\sigma_t$. 
If we standardize the (observed) returns $R_t$ by $\sigma_t$, that is, $R_t/\sigma_t\equiv \tilde{R_t}$, we expect to obtain the standard normal random time series $\epsilon_t$.
To test whether $\tilde{R_t}$ satisfies this expectation 
we need to estimate the volatility $\sigma_t$ of returns in the proper manner because volatility is not observable in the financial markets. 

In empirical finance, the standard technique to estimate volatility is to use volatility models such as 
GARCH-type models\cite{Engle1982autoregressive,Bollerslev1986JOE,Glosten1993JOF,Nelson1991Econ,Sentana1995RES,takaishi2017rational,takaishi2018volatility,bollerslev1992arch}.
The drawback of using such models is that volatility estimates are model-dependent, and which estimate is better is not well defined.
Recent availability of high-frequency financial data enable us to obtain a model-free estimate known as "realized volatility (RV)"\cite{andersen1998answering,mcaleer2008realized} 
constructed by the sum of the square of intraday returns.
The returns standardized by the RV are examined for the exchange rate\cite{andersen2000exchange,andersen2001exchange} and return\cite{andersen2001distribution}, 
and the normality of the standardized return $\tilde{R_t}$ is established. However, detailed studies report the existence of deviation from the normality caused by
the low sampling frequency in the RV\cite{andersen2007no,takaishi2012finite,takaishi2014realized,takaishi2016Nikkei,takaishi2018bias}.
Furthermore because $\tilde{R_t}$ is expected to be a random variable, we should observe no autocorrelation not only for $\tilde{R_t}$ but also for the absolute $\tilde{R_t}$, that is, $|\tilde{R_t}|$.
The autocorrelation of the absolute $\tilde{R_t}$ is examined for Japanese stocks\cite{takaishi2012analysis}, and it is verified that
no autocorrelation is observed for the absolute $\tilde{R_t}$\footnote{This behavior is also confirmed for returns simulated 
by the artificial spin model and standardized by GARCH volatility\cite{takaishi2013analysis}.}.
We examine the nonexistence of autocorrelation in the absolute $\tilde{R_t}$ of Bitcoin. 
Figure 3 shows the autocorrelation of the absolute $\tilde{R_t}$ standardized by the RV 
at a 5-min sampling frequency\footnote{In the empirical analysis, the 5-min sampling frequency for the RV calculations is 
optimal\cite{bandi2006separating,liu2015does}.},
and we confirm that there is no significant autocorrelation in the absolute $\tilde{R_t}$.

\begin{figure}
\centering
\includegraphics[height=6.5cm]{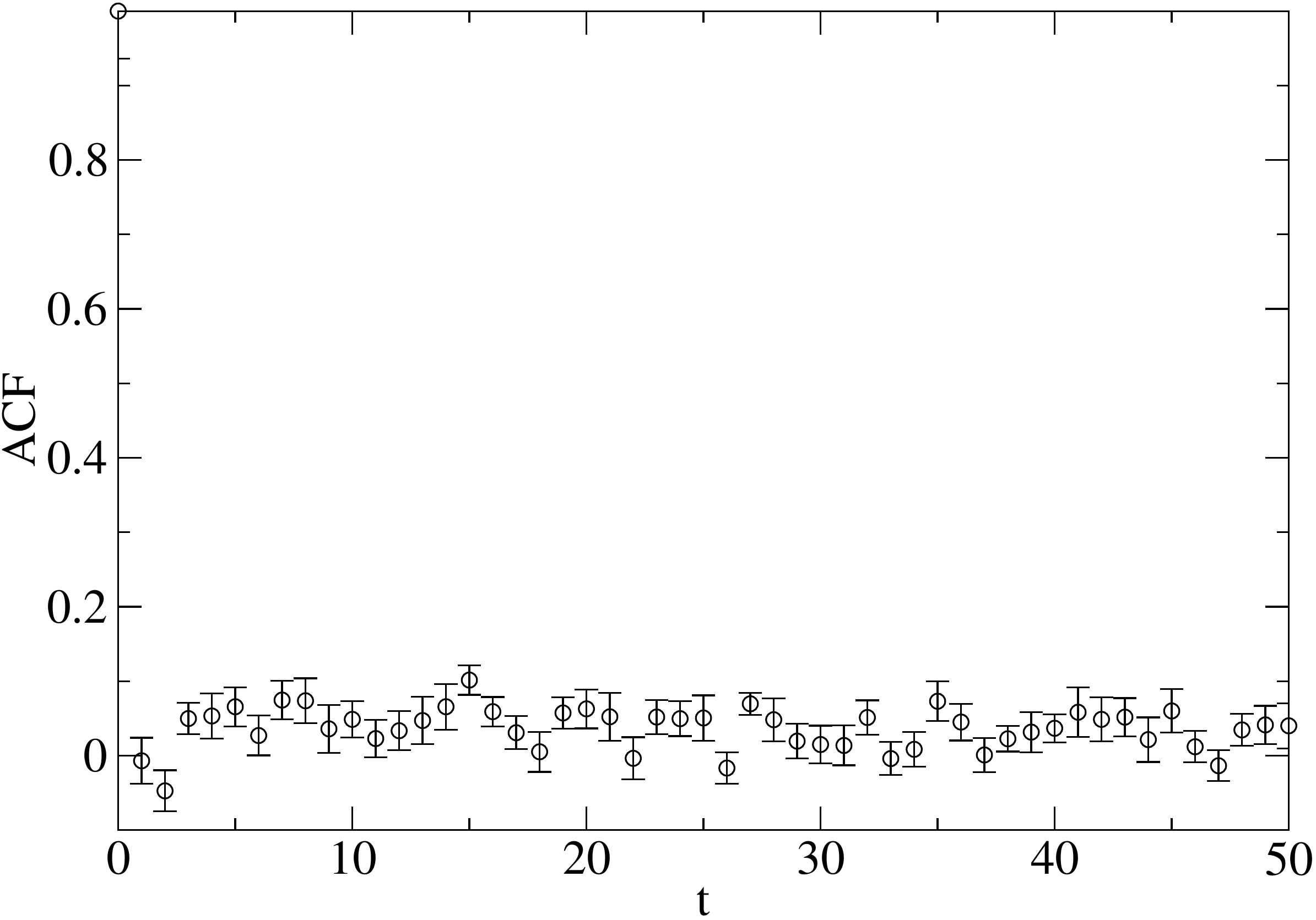}
\caption{
 ACF of the absolute returns standardized by RV.
For the calculation of the ACF, we take the data from January 10, 2015 to June 21, 2020.
}
\end{figure}

\section{Conclusions}
We investigate the scaling properties of the Bitcoin market 
and find that the tail index $\mu$ of the cumulative return distribution changes 
from $\mu \approx 2 $ to $\mu \approx 3$,
which indicates that the recent Bitcoin market exhibits
the well-established scaling law, that is, the inverse cubic law.
Our findings suggest that some properties of the Bitcoin market could change over time.

We find that the autocorrelation of absolute returns shows a power law with two scaling exponents
separated at approximately $t=1000$min.
The ACF of the absolute returns standardized by RV shows no autocorrelation,
which indicates that the return time series is consistent with a normal random variable with time-varying volatility.

\section*{Acknowledgment}
Numerical calculations for this work were carried out at the
Yukawa Institute Computer Facility and at the Institute of Statistical Mathematics.
This work was supported by JSPS KAKENHI Grant Number JP18K01556.

\section*{References}
%\bibliography{iopart-num}
\bibliography{mybibfile}
%\bibliography{icmsquare2020-1.bbl}

\end{document}